\input harvmac
\input epsf.tex
\def\N{{\cal N}}
\def\O{{\cal O}}
\def\G{{\cal G}}

\def\W{{\cal W}}
\def\F{{\cal F}}
\def\lfm#1{\medskip\noindent\item{#1}}

\lref\AEFJ{D.~Anselmi, J.~Erlich, D.~Z.~Freedman and A.~A.~Johansen,
``Positivity constraints on anomalies in supersymmetric gauge theories,''
Phys.\ Rev.\ D {\bf 57}, 7570 (1998)
[arXiv:hep-th/9711035].
}

\lref\AFGJ{D.~Anselmi, D.~Z.~Freedman, M.~T.~Grisaru and A.~A.~Johansen,
``Nonperturbative formulas for central functions of supersymmetric gauge  theories,''
Nucl.\ Phys.\ B {\bf 526}, 543 (1998)
[arXiv:hep-th/9708042].
}

\lref\Zam{A.~B.~Zamolodchikov,
``'Irreversibility' Of The Flux Of The Renormalization Group In A 2-D Field Theory,''
JETP Lett.\  {\bf 43}, 730 (1986)
[Pisma Zh.\ Eksp.\ Teor.\ Fiz.\  {\bf 43}, 565 (1986)].
}

\lref\NSVZ{V.~A.~Novikov, M.~A.~Shifman, A.~I.~Vainshtein and V.~I.~Zakharov,
``Exact Gell-Mann-Low Function Of Supersymmetric Yang-Mills Theories From Instanton Calculus,''
Nucl.\ Phys.\ B {\bf 229}, 381 (1983).
}

\lref\HO{H.~Osborn,
``N = 1 superconformal symmetry in four-dimensional quantum field theory,''
Annals Phys.\  {\bf 272}, 243 (1999)
[arXiv:hep-th/9808041].
}

\lref\GW{D.~J.~Gross and F.~Wilczek,
``Asymptotically Free Gauge Theories. 2,''
Phys.\ Rev.\ D {\bf 9}, 980 (1974).
}

\lref\BZ{T.~Banks and A.~Zaks,
``On The Phase Structure Of Vector - Like Gauge Theories With Massless Fermions,''
Nucl.\ Phys.\ B {\bf 196}, 189 (1982).
}

\lref\NSd{
N.~Seiberg,
``Electric - magnetic duality in supersymmetric nonAbelian
gauge theories,''Nucl.\ Phys.\ B {\bf 435}, 129
(1995)[arXiv:hep-th/9411149].
}

\lref\JE{J.~Erdmenger,
``Gravitational axial anomaly for four dimensional conformal field  theories,''
Nucl.\ Phys.\ B {\bf 562}, 315 (1999)
[arXiv:hep-th/9905176].
}

\lref\KINS{K.~A.~Intriligator and N.~Seiberg,
``Phases of N=1 supersymmetric gauge theories in four-dimensions,''
Nucl.\ Phys.\ B {\bf 431}, 551 (1994)
[arXiv:hep-th/9408155].
}
\lref\DKAS{
D.~Kutasov and A.~Schwimmer,
``On duality in supersymmetric Yang-Mills theory,''
Phys.\ Lett.\ B {\bf 354}, 315 (1995)
[arXiv:hep-th/9505004].
}

\lref\DKNSAS{D.~Kutasov, A.~Schwimmer and N.~Seiberg,
``Chiral Rings, Singularity Theory and Electric-Magnetic Duality,''
Nucl.\ Phys.\ B {\bf 459}, 455 (1996)
[arXiv:hep-th/9510222].
}

\lref\superspace{S.~J.~Gates, M.~T.~Grisaru, M.~Rocek and W.~Siegel,
``Superspace, Or One Thousand And One Lessons In Supersymmetry,''
Front.\ Phys.\  {\bf 58}, 1 (1983)
[arXiv:hep-th/0108200].
}

\lref\anselmi{D.~Anselmi,
``Inequalities for trace anomalies, length of the RG flow, distance  between 
the fixed points and irreversibility,''arXiv:hep-th/0210124.
}
\def\ev#1{\langle#1\rangle}
\
\def\r{{\bf r}}
\def\drawbox#1#2{\hrule height#2pt 
        \hbox{\vrule width#2pt height#1pt \kern#1pt \vrule width#2pt}
              \hrule height#2pt}

\def\Fund#1#2{\vcenter{\vbox{\drawbox{#1}{#2}}}}
\def\Asym#1#2{\vcenter{\vbox{\drawbox{#1}{#2}
              \kern-#2pt       
              \drawbox{#1}{#2}}}}
 
\def\fund{\Fund{6.5}{0.4}}

\Title{\vbox{\baselineskip12pt\hbox{hep-th/0304128}
\hbox{UCSD-PTH-03-02} }} 
{\vbox{\centerline{The Exact Superconformal R-Symmetry Maximizes $a$}}}
\centerline{ Ken Intriligator
and Brian Wecht}
\bigskip
\centerline{Department of Physics} \centerline{University of
California, San Diego} \centerline{La Jolla, CA 92093-0354, USA}

\bigskip
\noindent
An exact and general solution is presented for a previously open
problem.  We show that the superconformal R-symmetry of any 4d SCFT is
exactly and uniquely determined by a maximization principle: it is the
R-symmetry, among all possibilities, which (locally) maximizes the
combination of 't Hooft anomalies $a_{trial}(R)\equiv (9\Tr R^3-3\Tr
R)/32$.  The maximal value of $a_{trial}$ is then, by a result of
Anselmi et. al., the central charge $a$ of the SCFT.  Our $a_{trial}$
maximization principle almost immediately ensures that the central
charge $a$ decreases upon any RG flow, since relevant deformations
force $a_{trial}$ to be maximized over a subset of the previously
possible R-symmetries.  Using $a_{trial}$ maximization, we find the
exact superconformal R-symmetry (and thus the exact anomalous
dimensions of all chiral operators) in a variety of previously
mysterious 4d $\N =1$ SCFTs.  As a check, we verify that our exact
results reproduce the perturbative anomalous dimensions in all
perturbatively accessible RG fixed points.  Our result implies that
$\N =1$ SCFTs are algebraic: the exact scaling dimensions of all
chiral primary operators, and the central charges $a$ and $c$, are
always algebraic numbers.

\Date{April 2003}

\newsec{Introduction}

The 4d $\N =1$ superconformal algebra is $SU(2,2|1)$, whose bosonic
part is $SO(4,2)\times U(1)_R$.  Thus every $\N =1$ superconformal
field theory (SCFT) must have a conserved $U(1)_R$ symmetry, whose
current is in the same superconformal multiplet as the stress-energy
tensor.  There might be additional global flavor symmetries $\F$; the full
symmetry group of the $\N =1$ SCFT is then $SU(2,2|1)\otimes \F$. The
additional global symmetry $\F$ acts as a ``non-R'' symmetry (i.e.  the
supercharges are invariant). For example, $\N =1$ SQCD is
believed to flow to an interacting SCFT for $N_f$ in the range
$3N_c>N_f>{3\over 2}N_c$ \NSd, and the additional global symmetry of the
SCFT is $\F =SU(N_f)\times SU(N_f)\times U(1)_B$.  

The $U(1)_R$ symmetry residing in $SU(2,2|1)$ yields important and
{\it exact} results for SCFTs. 
For example, all operators $\O$ have dimension $\Delta (\cal O)$ which
satisfy
\eqn\deltar{\Delta(\O )\geq {3\over 2}R(\O) \qquad \hbox{and}
\qquad \Delta(\O )\geq -{3\over 2}R(\O),}
with $R(\O )$ the $U(1)_R$ charge of $\O$.  The first inequality is
saturated for pure chiral primary operators, and the second for pure
anti-chiral operators.  Since the $R$ charge of composite operators is
simply additive, so are the anomalous dimensions of composite chiral
primary operators; this is the statement that they form a chiral ring,
with non-singular mutual OPE.

The condition that the $U(1)_R$ global symmetry be free of ABJ type
anomalies, i.e. that $R$-charge conservation must not be violated in any
gauge field instanton backgrounds, is precisely the condition that the
NSVZ exact beta functions \NSVZ\ vanish for all gauge groups.

Another remarkable utility of the superconformal $U(1)_R$ symmetry
was found by Anselmi et. al. \refs{\AFGJ, \AEFJ}: the $U(1)_R$ 't Hooft
anomalies completely determine the $a$ and $c$ central charges of
the superconformal field theory:
\eqn\achooft{a={3\over 32}(3\Tr R^3-\Tr R), \qquad c={1\over 32}(9 \Tr R^3-
5\Tr R).}  Because of 't Hooft anomaly matching, this means that these
central charges can be computed simply in terms of the weakly coupled
UV spectrum, even for highly interacting IR fixed points.  It is
believed that the central charge $a$ obeys the 4d analog of
Zamolodchikov's c-theorem \Zam : Under any renormalization group flow,
perturbing away from any UV fixed point and flowing to a new IR fixed
point reduces the central charge: $a_{IR}<a_{UV}$.  This was verified
to be the case in many supersymmetric examples by using \achooft, as in
\refs{\AFGJ, \AEFJ}.  See \anselmi\ for some more recent developments.

So the superconformal R-symmetry is extremely useful...provided that
it can be found!   The symmetry constraints generally do not
uniquely determine $U(1)_R$ whenever  $\F$ is non-trivial.  This is because
if $R_0$ is some valid $U(1)_R$ symmetry, then so is 
\eqn\rtrial{R_t=R_0+\sum _I s_I F_I,}
where $F_I$ are all of the non-R flavor charges in the global symmetry group 
$\F$ \foot{ We emphasize
that all currents in \rtrial\ are bona fide symmetries, which are
anomaly free and respected by all superpotential terms. In particular,
we are not considering the situation discussed in e.g. \refs{\AFGJ,
\AEFJ}  of the RG flow of the $R$ current in the stress tensor
supermultiplet between its weak coupling expression and that of the
IR fixed point SCFT.  We are not considering RG flows here, only
aspects of the interacting SCFT RG fixed points.} and $s_I$ are arbitrary  
real parameters.  The superconformal $U(1)_R\subset SU(2,2|1)$ corresponds
then to some particular choice $\widehat s_I$ of these parameters, and we need
something beyond just symmetry considerations to determine what these are.  

There are some physical expectations which sometimes help.  For
example, the superconformal $U(1)_R$ is expected to commute with the
non-Abelian flavor symmetries, so we can restrict the linear
combination in \rtrial\ only to the Abelian flavor generators, which
commute with all non-Abelian elements of $\F$.  Also, if there's a
charge conjugation symmetry, the superconformal $U(1)_R$ should also
commute with that (e.g. for SQCD, baryons and anti-baryons should have the
same R-charge); then the $F_I$ in \rtrial\ can be restricted to those
commuting with charge conjugation.  For the case of SQCD, these constraints
imply that $U(1)_R$ can not mix with any of the generators of the global
flavor group $SU(N_f)\times SU(N_f)\times U(1)_B$; they thus 
uniquely determine the superconformal $U(1)_R$ of SQCD.  

Another physical requirement (which actually does not help to
determine the superconformal $U(1)_R$ symmetry) is that unitarity
implies that all gauge invariant, spinless operators must have
dimension $\Delta \geq 1$.  An operator $\Phi$ saturates this bound,
$\Delta (\Phi )=1$, iff it is a free field, with $\partial _\mu
\partial ^\mu \Phi =0$.  Thus all gauge invariant, chiral primary
operators must have $U(1)_R$ charge $R\geq {2\over 3}$, with the bound
saturated iff the operator is a free field.  But this unitarity
condition actually does not in any way constrain the R-charge
assignment of the basic fields entering the Lagrangian.  Indeed, it is
possible that the R-charge assignments of the basic fields is such
that some gauge invariant combinations appear to violate the $R\geq
2/3$ bound.  Examples where this happens appear in the theories
considered in \DKAS, where the resolution of the apparent conflict
with unitarity was also given: Any gauge invariant chiral operator,
$X$, which apparently violates the $R\geq 2/3$ condition is actually a
free field.  Because $X$ is decoupled from the rest of the SCFT,
there's an accidental extra $U(1)_X$ symmetry, under which only $X$ is
charged, which mixes with the $U(1)_R$ symmetry to correct the
superconformal R-charge of $X$ to be $R(X)=2/3$, with the R-charges of
the other operators unaffected.

In general, the superconformal R-symmetry is not uniquely
determined on symmetry grounds or the above considerations.
One well-known example is $\N =1$ SQCD with an added adjoint
and zero superpotential.  Because no other condition to determine the
superconformal R-symmetry had been previously known (as far as we are
aware), it had not been possible to apply the above powerful
constraints of superconformal invariance to generic $\N =1$ SCFTs.

We will here present and explore a simple prescription for uniquely
and exactly determining the exact superconformal $U(1)_R$ for any 4d
SCFT.  The idea is to parametrize the most general possible R-symmetry
as in \rtrial.  The subscript $t$ is for ``trial.''  The
superconformal $U(1)_R\subset SU(2,2|1)$ then corresponds to some
particular values, $\widehat s _I$,
\eqn\rspec{R = R_0 + \sum _I \widehat s_I F_I;}
our goal then is to determine the values of the $\widehat s_I$.  What
we show is that the $\widehat s_I$ can be uniquely determined by
imposing the following conditions on the 't Hooft anomalies:
\eqn\RRJ{9\Tr (R^2 F_I)=\Tr F_I,}
where $R$ is the $U(1)_R\subset SU(2,2|1)$ and $F_I$ are
all flavor charges in $\F$,  
and also 
\eqn\RJJ{\Tr R F_I F_K <0.} Specifically, this matrix has all
negative eigenvalues for all flavor symmetries.  
We will prove \RRJ\ and \RJJ\ using general properties of 4d $N=1$
SCFTs.   Plugging \rspec\ into \RRJ\ leads to a quadratic equation for 
each of the $\widehat s_I$.  Then \RJJ\ uniquely determines which are
the correct roots of the quadratic equations.  

The identities \RRJ\ and \RJJ\ have a nice interpretation.  Introduce 
\eqn\atrial{a_{trial}(s)\equiv {3\over 32}(3 \Tr R_t^3-\Tr R_t)}
for the general trial $R_t$ symmetry \rtrial\ as a function of the parameters
$s_I$.  When the $s_I$ are the special values $\widehat s_I$, where $R_t$
becomes the $U(1)_R\subset SU(2,2|1)$, the value of $a_{trial}$ is
the central charge $a$ of the SCFT, as in  \achooft.  The conditions \RRJ\ 
and \RJJ\ can be equivalently stated: the $U(1)_R
\subset SU(2,2|1)$ is precisely that which (locally) 
{\it maximizes} $a_{trial}(s)$!  This is simply because
\eqn\ranomid{{\partial a(s)\over \partial s_I}={3\over 32}
(9\Tr R_t^2F_I- \Tr F_I) \qquad\hbox{and}\qquad
{\partial ^2 a(s)\over \partial s_I\partial s_K}={27\over 16}~\Tr R_t F_IF_K.}
So \RRJ\ and  \RJJ\ together imply that, among all possible R-symmetries,
the $U(1)_R\subset SU(2,2|1)$ is that which (locally) maximizes $a$.

As an example of how $a_{trial}$ maximization determines the superconformal
$U(1)_R$, consider the case of a free theory
of $|\G |$ vector multiplets and $M$ chiral multiplets $\Phi _i$,
$i=1\dots M$, with trial charges $R_t(\Phi _i)=r_i$.  We then have
\eqn\rtrialex{a_{trial}={3\over 32}\left( 2|\G| +\sum _i 
[3(r_i-1)^3-(r_i-1)]\right).}
If we now extremize with respect to the $r_i$, we get
\eqn\riextr{(r_i-1)^2={1\over 9},}
with $r_i=2/3$ the root which is a local maximum and $r_i=4/3$ the
root which is a local minimum.  Our general identities \RRJ\ and \RJJ\
imply that the correct superconformal R-symmetry in this case is to
take $r_i=2/3$ as the charge of all chiral superfields.  This is indeed the
correct result for the free theory and, at this local maximum,
\eqn\afree{a=a_{free}={3\over 16}|\G |+{1\over 48}M.}  Note that, because 
$a_{trial}$ is a cubic function of the
$r_i$, there is no global maximum or minimum: $a_{trial}\rightarrow \pm \infty$
if we take $r_i \rightarrow \pm \infty$.

Several comments:
\lfm{1.} If a flavor symmetry $F_I$ has vanishing 't Hooft anomaly, 
$\Tr F_I=0$, then the condition \RRJ\ becomes $\Tr R^2F_I=0$.
Examples of such $F_I$ are non-Abelian flavor symmetries (which
satisfy $\Tr F_I=0$ by tracelessness of the generators) and $U(1)$
symmetries, such as baryon number, which do not commute with a charge
conjugation symmetry.  For such $F_I$ the solution of $\Tr R^2F_I=0$
is to take $R$ to commute with $F_I$.  So the condition \RRJ\
automatically ensures that the superconformal $U(1)_R$ does not mix
with non-Abelian flavor symmetries or baryon number, as was expected
on physical grounds.  For these non-Abelian or baryon number flavor
symmetries $F_I$, the remaining condition \RJJ\ becomes $(R(Q)-1)\Tr
F_IF_J<0$, where $R(Q)$ is the superconformal R-charge of those fields
$Q$, appearing in the Lagrangian, which are charged under such a
symmetry $F_I$.  
Here $R(Q)-1$ is the R-charge of the fermion component
of $Q$ and we used the fact that the superconformal $U(1)_R$ commutes
with these symmetries to take the $(R(Q)-1)$ outside the trace.
(This is for the case of a single irreducible representation
of the non-Abelian flavor symmetry with generators $F_I$.  With 
several such fields $Q_i$, in representations ${\bf r}_i$, 
we'd have $\sum _i (R(Q_i)-1)\Tr _{{\bf r}_i} F_IF_J<0$.)
Because all non-Abelian or baryon number flavor symmetries have
positive $\Tr F_IF_J$, we obtain a non-trivial new constraint on
SCFTs: all matter fields $Q$ in the Lagrangian which transform in
non-Abelian flavor representations, or which are charged under a
baryon number symmetry, must have superconformal R-charge $R(Q)<1$.
It is satisfying to verify that this new constraint is indeed
satisfied in all known examples of SCFTs, in all dual descriptions.

\lfm{2.}  The $a_{trial}$ maximization principle {\bf almost} 
immediately ensures the
$a$ theorem: $a_{IR}<a_{UV}$ for any RG flow between UV and IR fixed
points.  The reason is that generally $\F _{IR}\subset \F _{UV}$,
since the relevant deformations of the UV theory break some of the
flavor symmetries \foot{This also applies to Higgsings, if we
interpret $\F$ as also containing the global component of the gauge
group, which mixes with the flavor symmetries upon Higgsing.  The
previous comment ensures that $R$ commutes with the global component
of the gauge group too.}.  Since at the IR fixed point $a_{trial}$ is
maximized over a subspace of the parameter space of UV fixed point,
the maximal value will be smaller, showing that $a_{IR}<a_{UV}$.  The
``{\bf almost}" is because of two potential inadequacies in this
argument.  The first is that sometimes there are additional,
accidental flavor symmetries of the IR fixed point, so sometimes $\F
_{IR}$ isn't a subset of $\F _{UV}$.  The second caveat is that, since
the $a_{trial}$ maximum is only a local maximum, it's possible for the
maximal value on a restricted subspace to actually exceed that of the
larger parameter space, e.g. one could get a larger value than \afree\
if some added interaction restricts some of the R-charges to a
subspace where they're sufficiently large.  So the $a$ theorem
requires that those relevant deformations which can drive the theory
to a new RG fixed point can't possibly restrict the possible $R$
charges to a subspace where some are extremely large.  This fits with
the constraint mentioned in comment 1, and with the intuition that
interactions generally reduce the dimensions of chiral primary
operators.

\lfm{3.} The conditions \RRJ\ are quadratic equations for the 
$\widehat s_I$ appearing in \rspec.  The sign of the discriminants
of these equations must be such that the solutions $\widehat s_I$
are real, since the superconformal R-charge of all fields must be
real.  

\lref\ArgyresJJ{
P.~C.~Argyres and M.~R.~Douglas,
``New phenomena in SU(3) supersymmetric gauge theory,''
Nucl.\ Phys.\ B {\bf 448}, 93 (1995)
[arXiv:hep-th/9505062].
}
\lref\ArgyresXN{
P.~C.~Argyres, M.~Ronen Plesser, N.~Seiberg and E.~Witten,
``New N=2 Superconformal Field Theories in Four Dimensions,''
Nucl.\ Phys.\ B {\bf 461}, 71 (1996)
[arXiv:hep-th/9511154].
}

\lfm{4.} Our approach relies on being able to identify the full
symmetry group of the RG fixed point, e.g. via analyzing the UV
Lagrangian away from the RG fixed point.  But strongly interacting RG
fixed points can also have enhanced symmetries, which are not visible
in any weakly coupled Lagrangian description.  In particular, the
superconformal $U(1)_R$ of a SCFT could be such a symmetry.  This is
the case, for example, in the $\N =1$ and $\N =2$ SCFTs presented in
\refs{\ArgyresJJ, \ArgyresXN}.  Though our 't Hooft anomaly identities
should be applicable also at such RG fixed points, it remains to be
seen whether or not they can be used to determine the superconformal
$U(1)_R$ in such cases.

\lfm{5.} The superconformal $U(1)_R$ charges determined by our 
procedure outlined above will always be {\it algebraic} numbers,
i.e. rationals and roots of rationals.  This is because they are found
by solving quadratic equations with rational coefficients (which are
the 't Hooft anomalies).  Thus, for any SCFT, the exact anomalous
dimensions of all chiral primary operators, and the exact central
charges $a$ and $c$, will always be algebraic numbers.  (SCFTs of the
type discussed in the previous comment could be exceptions to this
general statement, though the known examples of this type actually
have rational R-charges).
\vskip .15cm

The outline of this paper is as follows.  In the next section we will
discuss some general aspects of currents, anomalies, and
supersymmetry.  After presenting some background material, 
we will argue for the result \RRJ\ by using a result due to Osborn \HO: 
the 3-point
function of two stress tensor supermultiplets and one flavor current
supermultiplet is of a form completely determined by the
superconformal symmetry with only a single overall multiplicative
coefficient to be determined.  
The condition \RJJ\ will also be obtained, by relating the 't Hooft
anomalies to current-current correlators and using unitarity, as in
\AEFJ.  In section 3 we verify, in complete generality, that our
$a_{trial}$ maximization precisely reproduces the known, leading
order, anomalous dimensions in all perturbatively accessible RG fixed
points of the type conjectured by \refs{\GW, \BZ}.  

In section 4 we use $a_{trial}$ maximization to obtain the exact
R-charge for some previously mysterious examples.  In particular, 
we consider in complete generality theories with two different 
matter field representations and zero superpotential.  A special 
case of this is $SU(N)$ with an adjoint chiral superfield and
$N_f$ fundamental flavors, with $W=0$.  This theory was argued to have 
a non-trivial RG fixed point in \KINS, and was further explored in 
\DKNSAS, but the exact R-charges could not be determined
before the present paper.  We also discuss some chiral quiver $\N=1$ 
SCFT examples,
both with and without superpotential, which gives another check of the
a-theorem.  

\newsec{Currents, anomalies, and supersymmetry}

In this section we argue for the result \RRJ\ by showing that
a result due to Osborn \HO\ implies that $\Tr R^2F_I$
and $\Tr F_I$ are necessarily proportional to each other.  
 As we also discuss, following
\AEFJ, the $\Tr RF_IF_J$ t'Hooft anomalies are proportional to the $\ev{
J^\mu _IJ^\nu_J}$ current correlator, whose sign is constrained by
unitarity.  The reader need not get too bogged down with the numerical
coefficients in the following section, since we only need to establish
general proportionality relations.  The constants of proportionality
can then always be determined by considering the particular case of a
free field theory.

\subsec{Review of currents and anomalies}

Let's first review some basics of currents and anomalies in theories
which aren't necessarily supersymmetric.  We'll call the gauge group
$\G$ and suppose, for illustrative purposes, that the flavor group is
a product $U(1)_1\times U(1)_2$, with left-handed chiral currents
$J^I_\mu$ for $I=1,2$.  (The generalization to non-Abelian flavor
symmetries is straightforward, with the most interesting aspects for
our discussions in the $U(1)$ factors anyway.)

The currents $J^I_\mu$ must not have ABJ anomalies, i.e. the triangle
diagram with a current insertion at one vertex and $\G $ gauge fields
at the other two must vanish.  Suppose that there are $n_i$ chiral
fermions $\psi ^i _\alpha$, all in $\G$ representation $\r _i$, with
$U(1)_I$ flavor charge $q^I_i$, which all run in the loop.  The vanishing
ABJ anomaly condition is   
\eqn\chargeaf{\sum _i q_i^I n_i \mu (\r _i)=0,}
where the $\mu (\r _i)$ are the
quadratic Casimirs from the coupling to the two $\G$ gauge fields,
\eqn\quadcasi{\Tr _{{\bf r}_i} (T_\G ^AT_\G ^B)=
\mu ({\bf r}_i)\delta ^{AB}, \qquad 2h \equiv \mu (Ad),}
and with $T _\G ^A$ the $\G$ generators in representation ${\bf r}_i$.
The anomaly free condition \chargeaf\ can
equivalently be stated as the condition that the fermion zero modes of
the $\G$ instanton's 't Hooft vertex must be flavor neutral.  The $\G$
instanton has $n_i \mu (\r _i)$ of the $\psi _{\alpha , i}$ fermion
zero modes.  (Our normalization is e.g. $\mu (\fund ) =1$ for $SU(N)$.)

We now consider current correlators.  In any conformal field theory,
the current two-point functions are completely 
determined by conformal invariance, up to the overall coefficient:
\eqn\JJ{\ev{J_\mu ^I(x)J_\nu ^K(0)}={\tau ^{IK} \over 16 \pi ^4}(\partial 
_\rho \partial ^\rho 
\delta _{\mu\nu}- \partial _\mu \partial _\nu)({1\over x^4}).}
In any unitary theory, $\tau _{IK}$ should be a matrix with all
positive definite eigenvalues.

The form of the current 3-point functions are also highly constrained.
The aspect of interest to us here is the anomalous violation of the current
conservation in contact terms. 
For example,
\eqn\jjjct{{\partial \over \partial z_\rho}
\ev{J^I_\mu (x)J^I_\nu (y) J^I_\rho (z)}=-{k_{III}
\over 48\pi ^2}\epsilon 
_{\mu \nu \kappa \sigma}{\partial \over \partial x ^\kappa}
{\partial \over \partial y ^\sigma}\delta (x-z)\delta (y-z),}
where the coefficient $k_{III}$ is the $\Tr U(1)_I ^3$'t Hooft anomaly 
\eqn\kanomis{k_{III}= \sum _i n_i |\r _i| (q_i^I)^3.}
Here $|\r _i|$ is the dimension of the representation $\r _i$.  

Similarly, there are current conservation
violating contact terms in current correlators involving mixtures of
the two currents, such as $\ev{J^L_\mu (x)J^L_\mu (y)J^I_\rho (z)}$ 
with $I\neq L$, which
has an anomalous contact term proportional to the mixed 't Hooft anomaly
\eqn\kmixed{k_{LLI}\equiv \Tr U(1)_L^2U(1)_I =\sum _i n_i |\r _i| 
q_i^I(q^L_i)^2.}
Another anomalous contact term, violating current conservation, occurs in the 
3-point functions involving the current at one vertex of the triangle diagram
and stress energy tensors at the other two, e.g. $\ev{T_{\mu \nu}(x)T_{\rho
\sigma}(y)J^I(z)}$, which has anomalous contact terms proportional to the
't Hooft anomaly
\eqn\kfis{k_I\equiv \Tr U(1)_I =\sum _i n_i |\r _i|q^I_i}

The contact terms in the above mentioned three point functions can be
conveniently expressed in terms of a lack of current conservation when
the currents $J^I_\mu$ are coupled to general background gauge fields
$A^I_\mu$, and the stress tensor $T_{\mu \nu}$ is coupled to a general
metric $g_{\mu \nu}$.  For example, we have
\eqn\curanom{{\partial \over \partial z_\mu } J^I_\mu
(z)={k_{III}\over 48\pi ^2} F_I \widetilde{F}_I +
{k_{IIL} \over 16\pi ^2} F_I \widetilde{F}_{L} +{k_{LLI}\over 16\pi
^2} F_L\widetilde{F}_{L}+{k_I\over 384\pi ^2} R \widetilde{R}} where
$L\neq I$, $F \widetilde{F} = {1 \over 2} \epsilon _{\mu \nu \rho \sigma}F^{\mu
\nu} F ^{\rho \sigma}$, and $R \widetilde{R} =  {1 \over 2}
\epsilon ^{\lambda \nu \rho \sigma}R_{\lambda
\nu \kappa \eta} R_{\,\,\,\,\,\,\, \rho \sigma}^{\kappa \eta}$.
 The first term comes from the descent formalism on
${1\over 3!}(F/2\pi i)^3$.  On the other hand, the third term comes
directly from the index theorem, as in the ABJ anomaly, involving
${1\over 2!}(F/2\pi i)^2$; this is one way to understand the relative
factor of three between these two terms (one can also see it directly
from the symmetry factors with the two corresponding triangle
diagrams).  The last term in \curanom\ is the Pontrjagin density; it
can be written in terms of the Weyl tensor (the Riemann tensor minus
all non-zero contractions of indices), 
so it vanishes in any conformally flat background.

\subsec{Supersymmetric Theories}

Consider a general 4d ${\cal N}=1$ supersymmetric gauge theory, with
gauge group $\G$, which we'll take to be simple to streamline the
present discussion (the generalization to product gauge groups is
easy).  The theory has chiral superfields $\Phi$ in the $\G$ representation
(which, of course, must be gauge anomaly free): $\oplus
_{i=1}^{s}n_i{\bf r_i}$.  If there is no added superpotential, the
full symmetry group of anomaly free global symmetries is
\eqn\flavgen{U(1)_R\times U(1)^{s-1}\prod _{i=1}^s SU(n_s).}  
(E.g. for
SQCD we have $s=2$, with ${\bf r_1}=\fund$, ${\bf r_2}=\overline{\fund}$,
and $n_1=n_2=N_f$.)

In the flavor group \flavgen, we have already eliminated one $U(1)$
classical global symmetry by the vanishing ABJ anomaly condition \chargeaf.
The chiral $U(1)_R$ symmetry in \flavgen\ assigns charge $1$ to the gauginos,
charge $R_i$ to the scalar components of the chiral superfields $\Phi _i$,
and charge $R_i-1$ to the fermion components of the chiral superfields;
thus the general condition \chargeaf\ of vanishing ABJ anomaly becomes
\eqn\ranomf{2h+\sum _i n_i (R_i-1)\mu (\r _i) =0.}
The remaining flavor symmetries in \flavgen\ are not R-symmetries, so
the gauginos are neutral and all components of the chiral superfield
$\Phi _i$, in representation $\r _i$, carry the same charge $q_i$.
According to \chargeaf, these must satisfy
\eqn\fanomf{\sum _i n_i q_i \mu ({\bf r}_i)=0.}
The non-Abelian part of the currents in \flavgen\ is always anomaly
free since their generators are traceless, so \fanomf\ only constrains
the overall $U(1)$ flavor symmetries in \flavgen, which commute with
the non-Abelian flavor symmetries. Any superpotential terms will further 
constrain the above flavor symmetries and charges, which could 
be easily incorporated into this discussion.

The anomalous contact terms in current three point functions or
equivalently the lack of current conservation when the global symmetries
are coupled to non-trivial backgrounds are as described in the previous
subsection, with the 't Hooft anomalies
\eqn\thooftg{\eqalign{\Tr ~R^3\equiv k_{RRR}&=|\G | + \sum _i n_i
(R_i-1)^3|\r _i|\cr
\Tr ~ R\equiv k_R&=|\G | + \sum _i n_i(R_i-1)|\r _i|\cr
\Tr ~R^2 F\equiv k_{RRF}&=\sum _i n_i(R_i-1)^2 q_i|\r _i|\cr
\Tr F^3\equiv k_{FFF}&=\sum _i n_iq_i^3 |\r _i|\cr
\Tr ~F\equiv k_F&=\sum _i n_iq_i |\r _i|\cr
\Tr ~RF^2\equiv k_{RFF}&=\sum _i n_i(R_i-1)q_i^2 |\r _i|.}}
$\G$ is the dimension of the gauge group, and $|\r _i|$ is that of $\r _i$.  

The flavor currents and their anomalies in general backgrounds can be
expressed in terms of current superfields.  One is the super-stress
tensor $T_{\alpha \dot \alpha}(x,\theta, \overline \theta)$, whose
$\theta =0$ component is the superconformal $U(1)_R$ symmetry,
components linear in $\theta _\beta$ and $\overline{\theta}_{\dot
\beta}$ are the supersymmetry currents, and terms quadratic in the
$\theta$ and $\overline \theta$ are the stress-energy tensor.  The
other, non-R, flavor currents reside in current superfields 
\eqn\flavsc{J_I(x,
\theta)= {1\over 4}\overline \Phi T_I \Phi, \quad\hbox{with current component}\quad
J_I ^\mu =\sigma ^\mu _{\alpha 
\dot \alpha} 
[\grad {}^\alpha, \grad {} ^{\dot \alpha}] J_I |_{\theta =0}.}
Here $\Phi$ and $\overline \Phi$
are the chiral and anti-chiral matter fields and $T_I$ is the appropriate
flavor generator, labeled by $I$.  
The super-stress tensor
$T_{\alpha \dot \alpha}$ couples to the metric in terms of the 
metric superfield $H^{\alpha \dot \alpha}$, while the non-R current superfields
\flavsc\ couple to background superfield vector multiplets $V_I$.  See 
\lref\BuchbinderQV{
I.~L.~Buchbinder and S.~M.~Kuzenko,
``Ideas And Methods Of Supersymmetry And Supergravity: Or A Walk Through  
Superspace,'' IOP Publishers (1998.  Bristol, UK).
}
\refs{\superspace,\BuchbinderQV} for background material and references. 

The super-stress tensor's anomaly can be expressed in the general
form \refs{\superspace,\BuchbinderQV}
\eqn\sstressw{\overline{\grad {}}^{\dot \alpha}T_{\alpha \dot \alpha}=
\grad {}_\alpha L_T,}
where $L_T$ is the trace anomaly, which can be written in terms of the
variation of the action with respect to the chiral compensator superfield of
supergravity. For example, in a theory with non-vanishing beta function we'd
have $L_T \sim \beta _1 
\tr W_g^2$, giving $T^\mu _\mu \sim \beta _1\Tr F_g^2$,
with $F_g$ the gauge field strength and $W_g$ its chiral superfield.
We're interested in conformal field theories, so $\beta =0$ and
$L_T=0$ when in flat space and trivial background gauge fields.  When
coupled to non-trivial backgrounds, however, we have
\eqn\ltis{L_T={c \over 24 \pi^2}\W ^2 -{a \over 24 \pi^2}\Xi 
-{1\over 96\pi ^2}\sum _{IK}\tau _{IK}\Tr W_I W_K.}
The coefficients $c$ and $a$ are the central charges,
$\W ^2 \equiv \half W_{\alpha \beta\gamma}W^{\alpha \beta\gamma}$ 
is the square of the super-Weyl tensor, and $\Xi \equiv \W ^2
+(\overline{\grad {}}^2+R)(G^2+2R\overline R)$ is the chirally projected
super Euler density; see appendix A of \AFGJ\ for a discussion of these
terms.  Taking components of the first two terms in \ltis\ 
is how \AFGJ\ obtained the relations \achooft.  The remaining terms in \ltis\
are the contributions to the scaling anomaly proportional to the super
field strengths $W_I$ of the background gauge fields $V_I$, which couple
to the flavor supercurrents $J_I$.  The coefficients $\tau _{IK}$ are the same
as those in \JJ, following the discussion in \AEFJ .

\subsec{The flavor super-anomaly}

We now turn to the anomaly of the non-R flavor currents $J_I$ in
general backgrounds.  The divergence of the current in \flavsc\ is
proportional to $D^2J -\overline D ^2 J$, so the super-anomaly
can be written as a non-zero chiral contribution to $\overline D^2J$.
We were not able
to find a discussion of this in the literature or supersymmetry
textbooks so, as far as we're aware, our result of $\overline D^2
J$ in a general background is new.  It's clear
how to proceed: since $\overline D^2 J$ is a chiral object, its non-zero
anomaly must involve chiral field strengths of the background fields.
Further, the current anomaly component of the result must reproduce the
terms in \curanom.  The result is 
\eqn\Janomis{\overline D ^2 J_I= {k_{III} \over 48 \pi^2} W _I^2+ 
\sum _{L \neq I}\left( {k_{IIL} \over 16 \pi^2}
W_IW_L + {k_{ILL} \over 16 \pi^2} W_LW_L\right)+{k_I \over 384\pi ^2}
\W ^2.}  (In a curved background, we replace 
\superspace\ $\overline D^2\rightarrow \widehat{\overline{\grad{}}}^2
\rightarrow \overline {\grad {}}^2+R$ in \Janomis.  We'll continue to
just call this $\overline D^2$; the implicit curvature term 
does not contribute to our final result.)

The crucial aspect of
\Janomis\ for our purposes is that the gravitational contribution only
involves $\W ^2$, with no additional term proportional to the chiral
Euler/Pontrjagin density $\Xi$.  This differs from the super-stress
tensor anomaly \ltis, which involves both chiral field strengths.  To
motivate the absence of an additional term proportional to $\Xi$ in
\Janomis, we note that $\Xi$ has a component which is the Euler
density, which is non-vanishing even for a conformally flat metric.
On the other hand, $\W ^2$ does vanish in a conformally flat background.
The absence of a term proportional to $\Xi$ in \Janomis\ is related
to the fact that supercurrents should be conserved in a conformally flat
background, since there is no
unavoidable charge violation by particle production in conformally flat 
backgrounds.  For example, in conformally flat AdS space,
with unbroken supersymmetry, the $J^\mu$ component of $J$ is conserved
because the Pontrjagin density vanishes.  Supersymmetry then implies
that the holomorphic quantity $\overline D^2 J$ identically vanishes.

The gravitational part of \Janomis\ can equivalently be described in
terms of an anomalous contribution to the current superfield 3-point
function \eqn\TTJanom{\overline D ^2_3 \ev{T_{\alpha \dot
\alpha}(1)T_{\beta \dot \beta}(2) J_I(3)}=\hbox{contact terms},} where
$(1)$ etc. label the superspace coordinates at the three points.  The
fact that there is only one independent gravitational term in
\Janomis, as opposed to the two terms in \ltis, is related to a result
of Osborn \HO\ for the relevant 3-point function:
\eqn\HOres{\ev{T_{\alpha \dot \alpha}(1)T_{\beta \dot \beta}(2)
J_I(3)}=K_I I_{\alpha \dot \alpha, \beta \dot \beta}(1,2,3).}  Here
$I_{\alpha \dot \alpha, \beta \dot \beta}(1,2,3)$ is a completely
determined function on superspace \HO, and the only dependence on the
theory is in a single overall coefficient $K_I$.  Taking $\overline{D} 
^2_3$ of
\HOres\ gives the contact terms indicated in \TTJanom.  This would
then lead\foot{Rather than
directly showing that the contact terms indeed lead to \Janomis, we
note that, because the function $I_{\alpha \dot \alpha, \beta \dot
\beta}(1,2,3)$ in \HOres\ is completely determined, the possible
gravitational contributions to $\overline D^2J$, namely $\W ^2$ and
$\Xi$, must appear with a fixed ratio.  Then a free-field calculation
would suffice to show that the coefficient of $\Xi$ is actually zero,
as in \Janomis.}  to \Janomis, with the single overall normalization constant
$K_I$, needed to fix the three-point function \HOres, thus
proportional to the $\Tr U(1)_I$ 't Hooft anomaly $k_I$.

In contrast, it was also shown in \HO\ that the $\ev{TTT}$ 3-point function
depends on two overall coefficients, which is related to the fact 
that both $\W ^2$ and $\Xi$ appear in \ltis.  
The fact that the 3-point function involving two stress tensors and
one flavor current $\ev{T_{\mu \nu}(1)T_{\rho
\sigma}(2)J_{I, \lambda}(3)}$ is completely fixed up to a single
overall normalization coefficient also holds in non-supersymmetric
conformal field theories, as shown by \JE.  There, too, we can say that
the single undetermined overall coefficient must be proportional to the
$\Tr U(1)_I$ 't Hooft anomaly.

\subsec{Why $a$ is maximized for the correct superconformal 
R-symmetry}

In this section we prove \RRJ\ and \RJJ, from which $a_{trial}$
maximization follows.

Consider \Janomis, with the background gauge fields $W_I$ set to 
zero and only the non-trivial backgrounds those in $\W^2$, namely
the metric and the $U(1)_R\subset SU(2,2|1)$
background gauge field strength, which we denote by $F_R$.  
The divergence of the $U(1)_I$ flavor current is 
\eqn\jidivi{\eqalign{\partial 
^\rho J^I_\rho &=-{i\over 4}[\grad {}^2 ,\overline{\grad{}^2}]J_I|_{\theta =0}=
{ik_I \over 384\pi ^2} (\grad {}^2\W ^2 -\overline{\grad{}}^2
\overline \W ^2)|_{\theta =0}\cr &={k_I \over 384\pi ^2} \left(R
\widetilde{R} + {8 \over 3} F_R \widetilde {F}_R \right).}}  Comparing with
\curanom, the first term is the expected result involving the anomaly
proportional to the $k_I=\Tr U(1)_I$ 't Hooft anomaly multiplied by
the Pontrjagin density.  The second term in \jidivi\ is the anomaly
which, as in \curanom, should be proportional to the $k_{IRR}=
\Tr U(1)_IU(1)_R^2$ 't Hooft
anomaly, but we see in \jidivi\ that it's instead also proportional to
$k_I$.  Taking into account the coefficients in \jidivi, as compared
with \curanom, we have thus shown the result \RRJ.  As a check of
eqns.  \Janomis\ and \RRJ, we note that \HOres\ suffices to show that
the ratio $k_I/k_{IRR}$ is some fixed number for all currents,
independent of the theory; so it can be evaluated for the case of free
fields, where the matter fermions have $R=-1/3$, showing that the
ratio is 9.  We can write the result \RRJ\ as
\eqn\RRJf{\sum _i n_i |\r_i| q_i \left(9(R_i-1)^2-1\right) =0.}

To prove \RJJ, suppose that we take trivial metric and $F_R$
backgrounds but turn on background gauge fields coupled to the
currents $J_I$.  Then \sstressw\ and \ltis\ imply, as in
\AEFJ,  that the $U(1)_R$ symmetry
has a divergence given by
\eqn\ridiv{\partial _\mu R^\mu = -{\tau _{IJ}\over 48\pi ^2} 
F_I \widetilde F_J, \quad \hbox{and hence}\quad k_{RIJ}\equiv
\Tr~ RF_IF_J=
-{\tau_{IJ} \over 3} ,} where $\tau _{IJ}$ are the same coefficients
as in \JJ.  Since $\tau _{IJ}$ must be a matrix with all positive
definite eigenvalues in any unitary theory, we see that supersymmetry,
along with unitarity, requires $\Tr~ RF_IF_J$ to be
negative, as in \RJJ. It is easy to see this in for example a free
field theory, where the R-charge of any matter fermion field is
$-1/3$.

\newsec{A check: comparing with perturbative RG fixed points.}

Let's write the beta function as:
\eqn\betais{\beta (g)=-\beta _1 g^3+\beta _2 g^5 +\dots.}
As pointed out in \refs{\GW,\BZ}, there can be a RG fixed point in the
perturbative regime if the one loop beta function is negative and the two
loop beta function positive, so $\beta _1$ and $\beta _2$ in \betais\
are both positive, and the coupling $g_*^2\approx \beta _1/\beta _2$ where
they cancel is sufficiently small.  The expectation is that, in this
case, higher order corrections might shift the fixed point value $g_*$
a bit, but not wipe out the qualitative feature of a RG fixed point.
This is the case when $\beta _1$ is very small, so the theory is just
barely asymptotically free.  In terms of a large $N_c$ expansion,
where $\beta _1$ is order $N_c$ and $\beta _2$ is order $N_c^2$
(corresponding to the expansion in 't Hooft coupling $g^2N_c$) we get
a small $\beta _1$ by adding matter flavors with total quadratic index
$\mu _T$ proportional to $N_c$, to make $\beta _1$ order $N_c^0$.  The
$\beta _2$ is still order $N_c^2$, so we get $g_*^2 N_c\sim 1/N_c$: the
't Hooft coupling is parametrically small, so the existence of the
RG fixed point is on fairly solid ground. 

We will be completely general, letting the gauge group be $\G$ and
there be matter chiral superfields in representations $\oplus _i
n_i \r _i$.  To be in the perturbative RG fixed point 
regime, we want to have
\eqn\betaiis{\beta _1 = 3h-\half \mu _T \equiv h \epsilon, \qquad 
\hbox{with}\qquad 0<\epsilon \ll 1,}
and $\mu _T\equiv \sum _i n_i \mu (\r _i)$ the total of the matter
field's quadratic Casimirs \quadcasi.  We first compute the
superconformal $U(1)_R$ symmetry via $a_{trial}$ maximization, and
then compare the result to a direct perturbative computation of the
anomalous dimensions at the RG fixed point.

We assign the fields R-charge $R_i$ subject to the vanishing anomaly
constraint \ranomf; to order $\epsilon$ we do this as
\eqn\rapprox{R_i = {2\over 3}+R_i^{(1)}\epsilon + \O (\epsilon ^2), 
\qquad \hbox{where}\qquad \sum _i n_i \mu (\r _i)R^{(1)}_i =-{2\over 3}h.}
We now compute
\eqn\rrjbz{9\Tr R^2F-\Tr F= -6\epsilon \sum _i R^{(1)}_in_i |\r _i| q_i
+\O (\epsilon ^2);} this must vanish for the superconformal $U(1)_R$.
This vanishing must hold for all possible choices of flavor charges
$q_i$ which satisfy the anomaly free condition \fanomf.
This requires that $R^{(1)}_i = \alpha \mu (\r _i)/|\r _i|$, and we can fix
the overall normalization $\alpha$ via \rapprox\ to obtain
\eqn\roneis{R_i={2\over 3}-\epsilon {2h\over 3}\left(\sum _j n_j 
\mu (\r _j)^2|\r _j|^{-1}\right)^{-1}{\mu (\r _i)\over |\r _i|}+
O(\epsilon ^2).}  

Of course, we could just as well obtain the
exact answer via $a_{trial}$ maximization, but the result will be
complicated, and we're only interested here in comparing the
order $\epsilon$ term to a perturbative computation.  

We now check that the result \roneis\ agrees with an explicit perturbative 
computation of the anomalous dimension, using
\eqn\dimi{R_i={2\over 3}\Delta _i = {2\over 3}(1+\half \gamma _i (g_*)),}
with $\gamma _i(g_*)$ the anomalous dimensions, evaluated at
the RG fixed point coupling $g_*$ where $\beta (g_*)=0$.  Working
to one-loop, which corresponds to order $\epsilon$, we have
\eqn\gammis{\gamma _i (g)=-{g^2\over 8\pi ^2} {|\G | \mu (\r _i) \over |\r_i|
}+O(g^4).}
This can be seen by considering the matter field propagator, with 
a single gluon (plus gluino) loop.  The group theory factor comes
{}from the sum over gluons
coupling to the matter field in representation $\r _i$: 
\eqn\TTid{
\sum _{a=1}^{|\G|}
T^a_{\r _i}T^a_{\r _i}={|\G |\mu (\r _i)\over |\r_i|}{\bf 1}_{|\r
 _i|\times |\r _i|},} with $T^a_{\r _i}$ the $\G$ generators in
 representation $\r _i$, with $a$ the adjoint index.  The group theory
 factors in \TTid\ are seen by comparing the trace of
\TTid\ with \quadcasi.   

The RG fixed point coupling $g_*$ is determined by requiring that
the NSVZ beta function vanish, which is equivalent to the condition
that the $U(1)_R$ symmetry \dimi\ satisfy the anomaly free condition 
\ranomf, i.e.
\eqn\gfp{2h+\sum _i n_i\mu (\r _i) (-{1\over 3}-{g_*^2\over 16\pi ^2}
|\G |\mu (\r _i) |\r _i|^{-1})+\O (g_*^4)=0,}
which gives
\eqn\ggpi{{g_*^2 |\G|\over 8\pi ^2}=(6h-\mu _T)\left(\sum _j n_i \mu (\r _j)^2
|\r _j|^{-1}\right)^{-1}.}
Recalling that $6h-\mu _T\equiv 2h\epsilon$, we see that
\dimi, together with \gammis\ and \ggpi\ indeed agree with 
\roneis.  It would be interesting to compare higher orders
in $\epsilon$, where our exact results make predictions about the
higher loop anomalous dimensions.

\newsec{Some previously mysterious examples}

\subsec{General case where $U(1)_R$ mixes with a single flavor $U(1)$.}

In this section, we will explicitly determine the superconformal $U(1)_R$ 
for cases where there is a single flavor current $J$ which can mix
non-trivially with $U(1)_R$.  This is the case, for example, for theories
with two types of representations (e.g. $SU(N)$ 
with $N_f$ fundamental flavors and an adjoint chiral superfield) and $W=0$.  

$a$ maximization leads to the $U(1)_R\subset SU(2,2|1)$ as given by
\eqn\rspecc{R=R_0+\widehat s J,}
where using \RRJ\ we have that $\widehat s$ is determined by 
\eqn\squad{\widehat s^2 \Tr J^3+2\widehat s 
\Tr R_0 J^2+ \Tr R_0^2 J -{1\over 9}\Tr J =0.}
This quadratic equation can be solved, with the correct root for
maximizing $a$, i.e. satisfying \RJJ, given by
\eqn\sis{\widehat s = { -\Tr R_0 J^2 - 
\sqrt{(\Tr R_0 J^2)^2-\Tr J^3(\Tr R_0^2J-{1\over 9}
\Tr J)} \over  \Tr J^3}.}
Of course $\widehat s$ must be real, so the quantity in the $\sqrt{}$ must be
positive in order for the theory to be superconformal.    

Let's apply this to a general theory with two kinds of matter representations
and $W=0$.  
Consider a theory with gauge group $\G$, $n_1$ matter fields in
representation $\r _1$ and $n_2$ matter fields in representation
$\r _2$.   We can take our initial $R_0$ to be one under which all chiral superfields
have the same charge, with value determined from \ranomf, and choose
our current $J$ to be one satisfying \fanomf.  Then the superconformal
R-symmetry is $R = R_0+
\widehat s J$: 
\eqn\rexg{R (\r _1)=1-{2h\over \mu _T}+{\widehat s \over n_1 \mu (\r _1)}, 
\qquad R(\r _2)= 1-{2h\over \mu _T}-{\widehat s\over n_2 \mu (\r _2)},}
where $\mu _T \equiv \sum n_i \mu (\r _i)$.  The value of  
$\widehat s$ is found by just plugging into \sis\ with 
\eqn\anomint{\eqalign{\Tr R_0 J^2&=-{2h\over \mu _T}\left(
{|\r _1|\over n_1\mu (\r _1)^2}+
{|\r _2|\over n_2 \mu (\r _2)^2}\right) \cr
\Tr R_0^2J-{1\over 9}\Tr J&=\left({4h^2\over \mu _T^2}
-{1\over 9}\right)\left( {|\r _1|\over \mu (\r _1)}-{|\r _2|\over \mu (\r _2)}\right) \cr
\Tr J^3&={|\r _1|\over n_1^2\mu (\r _1)^3}-{|\r _2|\over n_2^2 \mu (\r _2)^3}.}}
Asymptotic freedom, $\mu _T<6h$, is sufficient to ensure that $\widehat s$, 
as given by \sis , is indeed real.  

This general discussion can be applied, e.g. to the case
of $SU(N_c)$ with $N_f$ fundamental flavors and an adjoint 
chiral superfield.   We 
just  plug in $n_1=2N_f$ (counting fundamentals and anti-fundamentals together,
since they have the same R-charge, as in comment 1 in the introduction),
$|\r _1|=N_c$, $\mu (\r _1)=1$, $n_2=1$, $|\r _2|=N_c^2-1$, $\mu (\r _2)=2N_c$.
This yields 
$$R(Q)=R(\widetilde Q)={N_f\over N_c+N_f}+{\widehat s \over 2N_f}, \qquad 
R(\Phi)={N_f\over N_c+N_f}-{\widehat s \over 2N_c},$$
where $\widehat s$ is determined by \sis\ with 
$$\eqalign{\Tr R_0J^2&=-{N_c\over N_c+N_f}\left({N_c\over 2N_f}+{N_c^2-1\over 4N_c^2}
\right)\cr
\Tr R_0^2J-{1\over 9}\Tr J&=\left({N_c^2\over (N_c+N_f)^2}-{1\over 9}\right)\left({
N_c^2+1\over
2N_c}\right)\cr
\Tr J^3&={N_c\over 4N_f^2}-{N_c^2-1\over 8N_c^3}.}$$
As an example, we consider the case of $N_c\gg 1$, with fixed $N_f/N_c
\equiv \epsilon \ll 1$.
We then get, to leading order in $\epsilon$, $\widehat s \approx
2N_f(3-\sqrt{5})/3$, which gives
\eqn\exampchs{R(Q)=R(\widetilde Q)\approx {3-\sqrt{5}\over 3}, \qquad R(\Phi
)\approx {\sqrt{5}N_f\over 3N_c}.} We note that 
$R(Q)$ satisfies the $R<1$ constraint needed for fields transforming
non-trivially under non-Abelian flavor symmetries, as mentioned in
comment 1 of the introduction.  There are gauge invariant operators,
such as $Q\widetilde Q$ and $\Tr \Phi ^2$ which apparently violate the
$R\geq 2/3$ constraint but, as mentioned in the introduction, this just
means that these particular operators are actually free, and their
R-charge gets corrected to $2/3$ by additional accidental symmetries.
(This is similar to the situation when these theories are deformed by a $\Tr~
\Phi ^{k+1}$ superpotential \refs{\DKAS, \DKNSAS}.)    Note that \exampchs\ 
implies that the $\Tr~ \Phi ^{k+1}$ superpotential of \refs{\DKAS,
\DKNSAS}\ is a relevant deformation of the $W=0$ RG fixed point for
$k+1<6N_c/\sqrt{5}N_f$ in the above $N_c\gg 1$, $N_f\ll N_c$ limit.
This fits with the qualitative discussion of \DKNSAS\ on how this
superpotential can affect the IR physics, even when $k\geq 2$, despite
the fact that it naively appears to be irrelevant.  Also, as expected,
we find that $\Tr ~\Phi Q\widetilde Q$ is a relevant superpotential
deformation of the $W=0$ fixed point (driving the theory to ${\cal N}=2$
SQCD).  

\subsec{A quiver example and connection to AdS/CFT}

As an interesting example of a theory with more than one flavor
current, we consider the theory given by the quiver in Figure 1 with
gauge group $U(N)^4$. This quiver arises naturally in string theory as
the IR worldvolume theory of $N$ coincident
D3-branes placed at the tip of a noncompact Calabi-Yau  $X_6$, where
$X_6$ is a
complex cone over the first del Pezzo surface $dP_1$.  This string
theory construction leads to a non-zero superpotential, and the
large $N$ limit of that theory is an $\N =1$ SCFT which is dual to IIB
string theory on $AdS_5\times H_5$ with $H_5$ the horizon of the
complex cone over $dP^1$. This $H_5$ is a $U(1)$ fibration over $dP^1$.  We
will discuss this SCFT, for all $N$, shortly.

First, let's consider the theory with the quiver diagram of Fig 1 and with
no superpotential, $W=0$.  This theory has no known string theory construction
and is hence more mysterious than the theory with non-zero $W$.  We 
expect that the $W=0$ theory also flows to an interacting $N=1$ SCFT
in the IR, and we use the $a_{trial}$ maximization to determine the exact
$U(1)_R$ charges and hence the exact anomalous dimensions and 
central charges at this new RG fixed point.  
\bigskip
\centerline{\epsfxsize=0.30\hsize\epsfbox{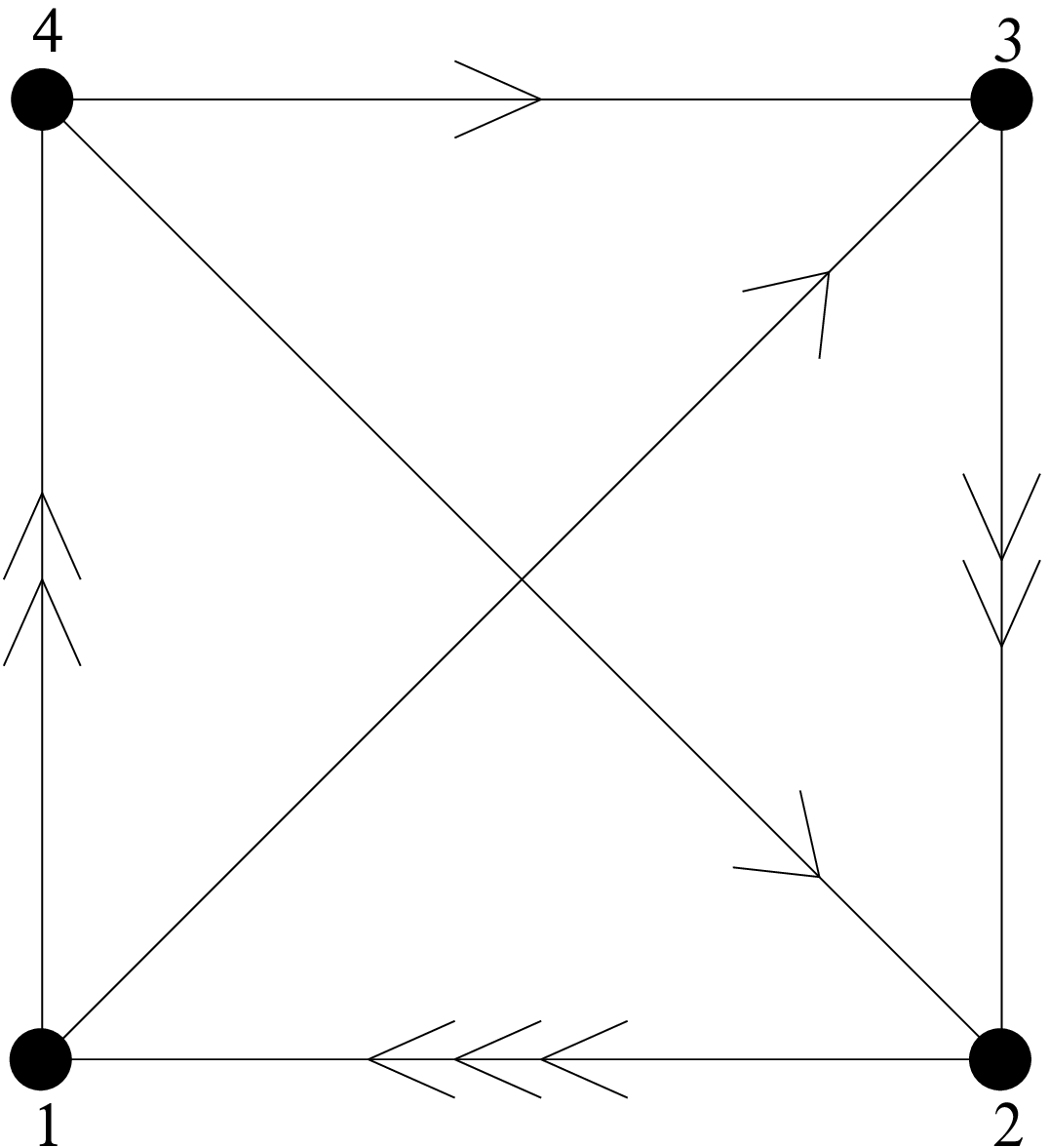}}
\centerline{\ninepoint\sl \baselineskip=8pt {\bf Figure 1:}
{\sl The $U(N)^4$ quiver diagram.}}
\bigskip

The IR SCFT has symmetry group $SU(2,2|1)\times \F$, with flavor group
$\F = U(1)_1\times U(1)_2 \times SU(2)\times SU(2)\times SU(3)$.
The non-Abelian symmetries rotate the multiple bi-fundamental flavors
and classically there's an additional flavor $U(1)^6$, 
one for each of the six legs of the quiver.  Enforcing the anomaly free 
condition for each of the four gauge
groups reduces this $U(1)^6$ down to $U(1)^2$.  As
discussed in the introduction, $a_{trial}$ maximization implies that
only the $U(1)^2$ can mix non-trivially with $U(1)_R$.  Without loss
of generality, we can assign $U(1)$ charges as:
\eqn\chtab{ \matrix{ &J_1&J_2&R_0 \cr X_{21}& -2/3& -1/3& 3/4 \cr
X_{14}& 1& 0&1/2 \cr X_{43}& -2& -1& 1/4 \cr X_{32}& 1& 0& 1/2 \cr
X_{13}& 0&1&3/4 \cr X_{42}&0&1& 3/4. \cr}} These charges form a basis
of the solutions of the appropriate anomaly-free conditions at each
node.  (Of course, any non-degenerate linear combinations of these
would be an equally valid basis choice.)  For completeness we include
all relevant 't Hooft anomalies here (omitting an overall factor of
$N^2$ for each): $\Tr R_0 = 0, \Tr R_0^3 = 3, \Tr J_1=\Tr J_2=0,
\Tr J_1^3 = -{44 \over
9}, \Tr J_2^3 = {8 \over 9}, \Tr J_1^2 J_2 = -{40 \over 9}, \Tr J_1
J_2^2 = -{20 \over 9}, \Tr R_0^2 J_1 = -{1 \over 4}, \Tr R_0^2J_2 =
-{1 \over 2}, \Tr R_0J_1^2 = -{16 \over 3}, \Tr R_0J_2^2 = -{4 \over
3}, \Tr R_0J_1 J_2 = -{5 \over 3}.$

We write the exact superconformal $U(1)_R$ symmetry as $R=R_0+
\widehat s_1 J_1+\widehat s_2 J_2$, and determine the values
of $\widehat s_I$ by imposing $9\Tr R^2J_I=\Tr J_I$, for $I=1,2$, and
use \RJJ\ to determine the correct roots of the resulting quadratic
equations; the result is $\widehat s_1={-2+\sqrt{5}
\over 6}$ and 
$\widehat s_2={-7 + 2\sqrt{5} \over 12}$.  We thus obtain for the
exact R-charges
\eqn\rchar{\eqalign{R(X_{21})&={7-\sqrt{5}\over 6}, \quad R(X_{14})=R(X_{32})=
{1+\sqrt{5}\over 6},\cr R(X_{43})&={3-\sqrt{5}\over 2}, \quad
R(X_{13})=R(X_{42})={1+\sqrt{5}\over 6}.}}  Since the $R$ charges of
all gauge invariant chiral operators, such as $X_{21}X_{14}X_{42}$,
satisfy the condition $R>2/3$, these operators are not free fields.
The value of the central charge $a$, found by plugging \rchar\ into
\achooft\ is
\eqn\aexis{a={3+5\sqrt{5}\over 16}N^2 \approx 0.886 N^2.}

One final comment on our proposed RG fixed point for the $W=0$ theory.
Some of the nodes of the quiver have $N_f=3N_c$, and are thus not
asymptotically free (the one loop beta function vanishes, and the two
loop beta function is positive).  Their gauge couplings would flow to zero
in the IR, in the absence of other
interactions.  Our claim is that, upon including the gauge interactions of
the other nodes, the theory flows to an IR fixed point SCFT, where all
gauge groups are interacting.  The NSVZ beta function for each gauge
group vanishes, since \rchar\ is anomaly free.  
We have not proven that the fixed point exists, since we haven't proven that
the dynamics can actually realize the R-charges \rchar, but the various 
consistency checks give us confidence that
the fixed point SCFT exists.

Let's now consider the theory with the non-trivial superpotential, as obtained
via $D3$ branes at the singularity of a local CY which is a complex cone 
over $dP1$:
\eqn\supi{W=X_{21}X_{14}X_{42} +
X_{21}X_{13}X_{32} + X_{21}X_{14}X_{43}X_{32}.}  Since we only care
about the symmetries, we've suppressed the coefficients and flavor
indices; see 
\ref\FengZW{
B.~Feng, S.~Franco, A.~Hanany and Y.~H.~He,
``Symmetries of toric duality,''
JHEP {\bf 0212}, 076 (2002)
[arXiv:hep-th/0205144].
} for the precise superpotential.  We can think of the resulting SCFT as
the IR limit of a RG flow, where the UV limit of the RG flow 
is the above described $W=0$ SCFT, deformed by the relevant 
perturbation \supi.  We now obtain the exact R-charges of this
IR SCFT.

The superpotential \supi\ reduces the $U(1)^2$ flavor symmetry of the
$W=0$ theory to $U(1)$: the superpotential respects the $R_0$ symmetry
in \chtab\ (since all terms in the superpotential have R-charge 2) but
is only neutral under $J=-2J_1+J_2$.  So rather than maximizing
$a_{trial}$ with respect to $s_1$ and $s_2$ independently, as before,
we now maximize subject to the constraint that $s_1=-2s_2\equiv s$.
We can impose \RRJ\ by simply plugging into \sis, which gives
$\widehat s=0$.  Thus the $R_0$ charges in \chtab\ are the exact
R-charges for the theory with superpotential \supi.  Using these
charges, one finds $a=27N^2/32 \approx 0.844N^2$. This is less than
the value \aexis\ which we found in the $W=0$ case, which is
consistent with the a-theorem.

String theory gives another way to determine the exact central charge
$a$, and also the exact R-charges, in terms of the $H_5$ geometry.
In particular, the baryonic operators such as $B_{12}=\det X_{12}$
correspond to particles in $AdS_5$, which arise from D3 branes wrapped
on 3-cycles of $H_5$.    
The R-charges of the baryons, and hence the bi-fundamentals,
are then related to the volume of the $H_5$ 3-cycles which the D3 wraps.  
This can be computed and shown to agree perfectly with the above results
\lref\IntriligatorWR{
K.~Intriligator and B.~Wecht,
``Baryon charges in 4D superconformal field theories and their AdS  duals,''
arXiv:hep-th/0305046.
}
\IntriligatorWR.

\vskip 1cm

\centerline{\bf Acknowledgements}

We would like to thank T. Banks, M. Gross, J. Kumar, D. Kutasov,
H. Osborn, M. Rocek, I. Rothstein, and C. Vafa for discussions and
correspondence.  This work was supported by DOE-FG03-97ER40546.

\listrefs
\end